\documentclass[10pt,twocolumn,preprintnumbers,amsmath,amssymb,nofootinbib,superscriptaddress]{revtex4-1}

\usepackage{amsmath}  
\usepackage{amssymb}  
\usepackage{theorem}

\usepackage{graphicx}
\usepackage{dcolumn}
\usepackage{amssymb,amsmath,bm}
\usepackage{color}
\usepackage[colorlinks,linkcolor=red,citecolor=blue,urlcolor=blue ]{hyperref}
\usepackage{multirow}
\usepackage{enumitem}
\usepackage{mathptmx} 
\usepackage{tensor}
\usepackage{amssymb} 
\usepackage{lettrine}
\usepackage[normalem]{ulem}
\usepackage{empheq}
\usepackage{comment}
\usepackage{subcaption}

\usepackage{booktabs}

\newcommand{\mnras}{Mon.\ Not.\ R.\ Astron.\ Soc.\ }

\newcommand{\ssr}{Space\ Sci.\ Rev.\ }


\begin{document}
  \title{Strong-lensing rates of massive black hole binaries in LISA}
  
\author{Juan Gutiérrez}
 \email{j.gutirrezmunizaga@uandresbello.edu}
  \affiliation{Instituto de Astrofísica, Departamento de Física y Astronomía, Universidad Andrés Bello, Santiago, Chile}
   \author{Macarena Lagos}
 \email{macarena.lagos.u@unab.cl}
  \affiliation{Instituto de Astrofísica, Departamento de Física y Astronomía, Universidad Andrés Bello, Santiago, Chile}

\begin{abstract}
Similarly to electromagnetic (EM) signals, gravitational lensing by intervening galaxies can also affect gravitational waves (GWs). In this paper, we estimate the strong-lensing rate of massive black hole mergers observed with LISA. Given the uncertainties in the source populations as well as in the population of galaxies at high redshift, we consider: six different source population models, including light and heavy seeds, as well as three lens population models, including redshift-independent and redshift-dependent evolution properties. Among all the scenarios explored, the expected number of strong lensed events detected in a 4-year observation time in LISA ranges between 0.13-231 with most of them having two (one) images detectable in the heavy (light) seed scenarios. The event numbers obtained correspond to $0.2\%-0.9\%$ of all detected unlensed events. Out of all the detectable strong-lensed events, up to $61\%$ (in the light-seed scenario) and $1\%$ (in the heavy-seed scenario) of them are above the detectability threshold solely due to strong lensing effects and would otherwise be undetectable. For detectable pairs of strong-lensed events by galaxy lenses, we also find between $72\%-81\%$ of them to have time delays from 1 week to 1 year.
\end{abstract}


\maketitle

\section{Introduction}\label{sec:intro}

Gravitational waves (GWs) offer a unique opportunity to probe the universe, as they travel largely unimpeded over cosmological distances. In particular, with nearly 100 detections by LIGO-Virgo-KAGRA (LVK) collaboration \cite{KAGRA:2021vkt} and thousands more expected in coming years \cite{KAGRA:2013rdx}, GWs provide an untapped resource for dark matter studies. 

Similarly to electromagnetic (EM) signals, gravitational lensing by intervening dark matter halos can magnify GWs, potentially allowing detection at even higher redshifts. If the source, lens and observer are highly aligned, signals suffer from strong-lensing effects, in which multiple (de)magnified and time-delayed images from the same source are received by the observer \cite{1992grle.book.....S}. In the case of GWs, the joint analysis of such multiple images is expected to considerably improve the estimation of source sky localization \cite{Hannuksela:2020xor} and open the possibility of multi-messenger studies with EM follow-ups (see recent reviews in \cite{Birrer:2025rdy, Smith:2025axx}). Applications of strong-lensed GW events include cosmography measurements \cite{Liao:2017ioi, Li:2019rns, Cao:2019kgn}, tests of gravity and dark matter tests \cite{Fan:2016swi, Diego:2019rzc, Goyal:2020bkm, Ezquiaga:2020dao, Ezquiaga:2022nak, Chung:2021rcu, Narola:2023viz, Wright:2024mco, Takeda:2024ghe}. In addition, the long wavelength of GWs can further suffer from wave-optics lensing effects whenever the characteristic size of the lens becomes comparable to or larger than the GW wavelength. In the case of LISA, these effects are expected to become relevant when the lens is a low-mass dark matter halo, with characteristic lens masses $\lesssim 10^8M_\odot$ (and associated virial masses of order $\lesssim 10^{10}M_\odot$) \cite{Takahashi:2003ix}. These wave-optics effects have been studied in LISA \cite{Takahashi:2003ix, Caliskan:2022hbu, Brando:2024inp,Pijnenburg:2024btj, Singh:2025uvp}, although they will not be considered here, and hence we will focus on lenses with masses over $10^8M_\odot$. 

Furthermore, we emphasize that lensing is an inevitable cosmological effect and if not accounted for, it can lead to biases in inferred binary parameters and population studies. On the one hand, weak and strong lensing would affect the inferred distance and source black hole mass \cite{Hirata:2010ba, Canevarolo:2023dkh, Mpetha:2024xiu}, with strong lensing additionally changing the inferred number of sources in a population study as well as possibly inducing phase changes confused with modified gravity \cite{Ezquiaga:2022nak}. On the other hand, wave optics induces frequency-dependent distortions in the phase and amplitude that could potentially bias various binary parameters \cite{Shan:2023qvd}, and even create biases towards modified gravity \cite{Ezquiaga:2020spg, Mishra:2023vzo, Liu:2024xxn}.

For current ground-based GW detectors, a search of lensing on GW events was performed in \cite{LIGOScientific:2021izm, LIGOScientific:2023bwz}, showing no evidence for strong lensing to date. Nonetheless, in the next years the first GW strong lensed events are expected \cite{Ng:2017yiu, Wierda:2021upe}. Furthermore, with the advent of next-generation GW detectors, such as Cosmic Explorer \cite{Evans:2021gyd} and Einstein Telescope \cite{Abac:2025saz}, lensed GW events will become commonplace. About a hundred of strongly gravitationally lensed GW events could be detected per year in the next decade \cite{Li:2018prc, Wang:2021kzt}. 

For space-based GW detectors, the Laser Interferometer Space Antenna (LISA) \cite{LISA:2024hlh} is set to begin construction in 2025, and will detect GWs from massive black hole binaries (MBHBs) up to redshift $z\sim 20$. While the population properties and detection rates of MBHBs are not entirely known, the high-redshift reach of LISA suggest promising lensed events. Forecasts on wave optics lensing caused by low-mass dark matter halos have been performed, with a detection of up to 8 events in 4 years \cite{Caliskan:2023zqm}. In addition, initial work on estimating the strong lensing rate of LISA was performed in \cite{Sereno:2010dr}, where it was concluded that a few strong lensed events would be detectable in a 5-year mission. %\ml{add more about EMRI lensing studies as well}

In this paper, we update and build up on the estimations performed in \cite{Sereno:2010dr} over a decade ago. In particular, we consider the latest simulations of MBHBs in \cite{Barausse:2012fy, Klein:2015hvg, Antonini:2015sza, Barausse:2020mdt, Barausse:2023yrx}, including heavy and light seeds as initial mass functions at high redshift. Due to the unknown properties and evolutions of MBHBs, we consider six different source population models, which are found to lead between 73-37,979 unlensed detectable events in a 4-year observation window time. 
We calculate the strong-lensed event rate in a 4-year mission and we analyze how many events would have single or double images detectable. Since LISA sources can be detectable up to redshift $z\sim 20$ and galaxy population properties are not known at such high redshifts, we consider three different galaxy lens population models with different redshift evolutions. Overall, we find the strong-lensing detectable event rate to range between 0.13-231 events in 4 years. Note that this range is wider, and the upper limit much higher than the one obtained previously in \cite{Sereno:2010dr}. Most of this uncertainty comes from the source population models, whereas the different galaxy lens models cause variations of about 40\% in the calculated rates. All the results are available in detail in \cite{Github_JG_2025}.

As shown in previous studies \cite{Haris:2018vmn, Li:2018prc, Yu:2020agu, LIGOScientific:2021izm, LIGOScientific:2023bwz}, the incorporation of priors on time-delay distributions of images can be an important ingredient to confidently identify strong-lensed events. For this reason, here we also include an analysis on time delays of the strong-lensed events expected in LISA. Here we find between 72\%-81\% of the detectable strong-lensed events to have time delays between 1 week to 1 year, depending on the source and lens population model.

This paper is organized as follows. In Sec.\ \ref{sec:lensing} we revisit the strong lensing framework, including the models of MBHBs sources and galaxy lens populations used. In Sec.\ \ref{sec:unlensed} we describe the unlensed event rates and their expected signal-to-noise ratio (SNR) in 4 years, for the different MBHBs source models. In Sec.\ \ref{sec:rates} we present the main results regarding the statistical properties of strong-lensed events, including their detection rates, SNRs, and time delay distributions. Finally, in Sec.\ \ref{sec:discussion} we summarize our results and discuss their future impact. Whenever necessary, cosmological quantities will be calculated using the best-fit Planck 2018 cosmological parameters \cite{Planck:2018vyg}.

\section{Framework}\label{sec:lensing}
\begin{figure}[h!]
    \centering
    \includegraphics[width=1.0\linewidth]{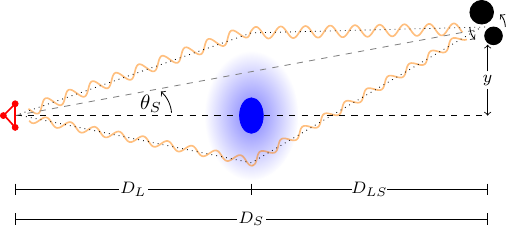}
    \caption{Lensing configuration of a MBHB source located at an angular diameter distance $D_S$ from the LISA detector, with a dimensionless impact parameter $y$, which suffers from strong lensing by a galaxy at an angular diameter distance $D_L$ from the detector. The signal can take two trajectories from the source to the detector, as indicated by the dotted lines, which results in two GW signals being detected with different amplitudes and arrival times.}
    \label{fig:SLsetup}
\end{figure}

In this paper, we will analyze a lensing scenario illustrated in Fig.\ \ref{fig:SLsetup}. A binary black hole is the source, which is located at an angular diameter distances $D_S$ from the observer. The source is displaced by an angle $\theta_s$ and quantified by a dimensionless impact parameter $y$ with respect to the optical axis (observer-lens line). This parameter will determine if the system is in the regime of weak or strong lensing. In addition, the lens will be considered to be typically a galaxy, located at an angular diameter distance $D_L$ from the observer. Multiple images can be observed from a single source, each one taking a different trajectory and probing a different region of the lens gravitational potential. As a result, each image suffers from a different time delay and magnification. In the case of GWs from binary systems, the signal has a coherent phase and the multiple images will also have a different constant phase shift \cite{Dai:2017huk, Ezquiaga:2020gdt}. Nonetheless, for the purpose of estimating lensing rates, these phase shifts will not be relevant and will be ignored here.

\subsection{Source Population}\label{sec:sources}
Since there is still uncertainty as to what is the formation channel of MBHs in the universe, results from semi-analytic simulations of different formation channels are considered, following the improvements of the original work in \cite{Barausse:2012fy} denoted as K+16 \cite{Klein:2015hvg, Antonini:2015sza}, and B+20 \cite{Barausse:2020mdt, Barausse:2023yrx}. 

We will consider two high-redshift seeds leading to MBHBs in quasi-circular orbits: a light-seed model (denoted as LS/popIII), where massive BHs are assumed to form at high redshift $(z>15-20)$ in the mass range $10^2-10^3 M_\odot$ from the collapse of heavy Pop III stars in the most metal-poor dark matter halos; and a heavy-seed model in which most of the mass in a protogalactic disk collapses into a supermassive star or a quasi-star leaving behind a BH in the mass range $10^4-10^5 M_\odot$ at redshifts $z\sim 8-15$ (denoted as HS/Q3). 

Various specific models for both seeds are available, which mainly differ in the time delay considered between the MBHB merger and their host galaxy mergers. As previously concluded in \cite{EPTA:2023xxk, Barausse:2023yrx} recent Pulsar Timing Array (PTA) observations, large delays at separations of hundreds of pc are disfavored and MBHBs merge efficiently after galaxy mergers. Therefore, we will only consider models with no delay or medium time delays in agreement with PTA data: (i) “Q3-nod (K+16)”, “HS-nod-noSN (B+20)”, “HS-nod-SN (B+20)” and “HS-nod-SN-high-accr (B+20)” which assume MBHs binaries to form with no delay from a galaxy merger; (ii) ``PopIII-d (K+16)” and “Q3-d (K+16)” which account for some delays between the MBHB and the galaxy merger such as stellar hardening, MBH triple interactions, and planetary-like migration. %\ml{should we add the LS-nod-SN and LS-nod-nSN?}
Here, the labels “SN” and “noSN” indicate models that do and do not account for the effect of supernova (SN) feedback on the growth of nuclear gas reservoir, respectively. Models at finite and extrapolated infinite resolution are publicly available\footnote{ \url{https://people.sissa.it/~barausse/catalogs/}}. Here we use the infinite resolution simulations, except for the HS-nod-SN-high-accr (B+20) model, whose finite resolution predictions agree with PTA measurements \cite{Barausse:2023yrx}.
Notice that the previous LISA strong lensing estimates in \cite{Sereno:2010dr} used older source population simulations developed in \cite{Volonteri:2002vz, 2006MNRAS.370..289B, 2008MNRAS.383.1079V}.

\begin{figure}[h!]
    \centering
\includegraphics[width=\linewidth]{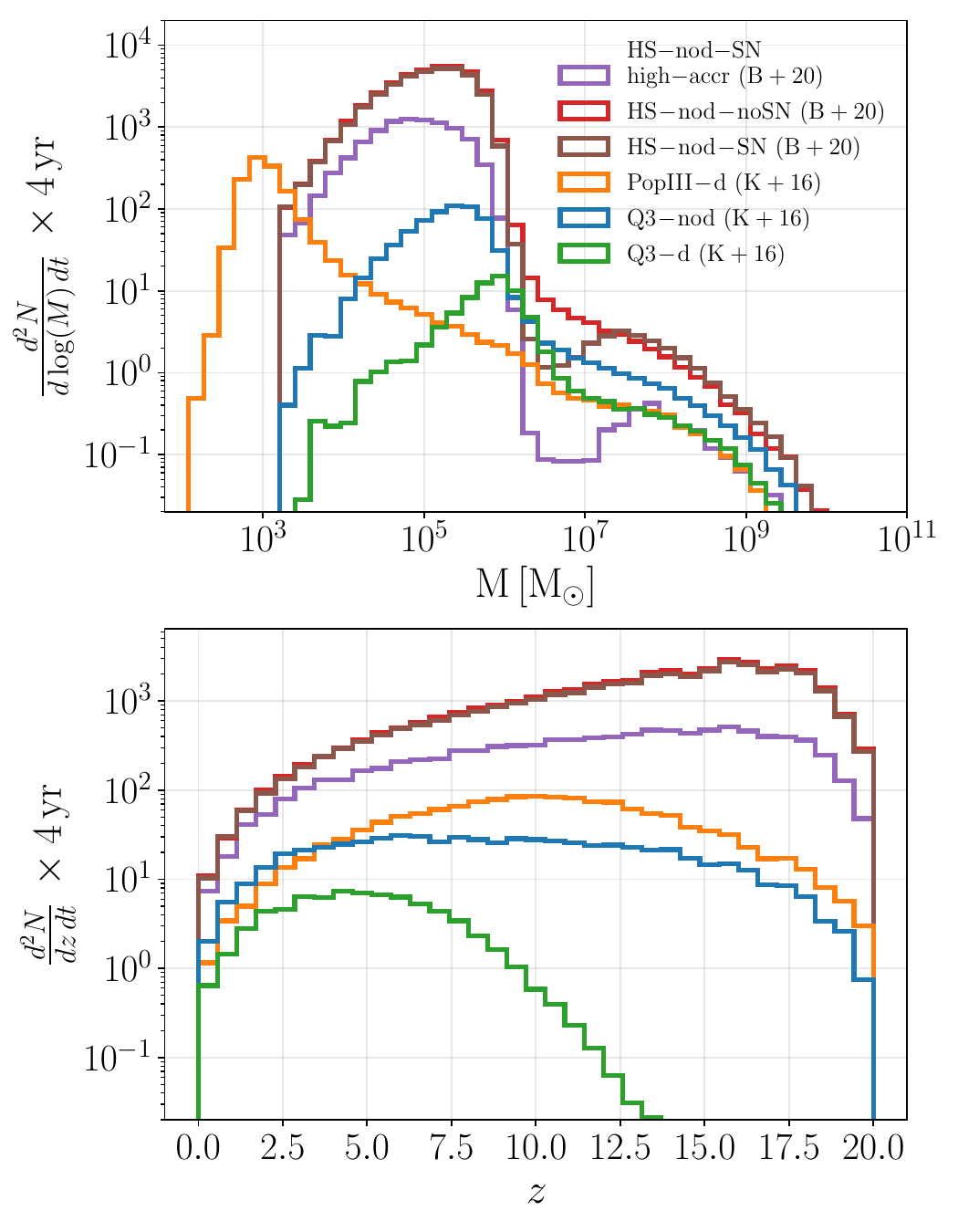}
\caption{Intrinsic merger rate distribution of MBHBs in 4 years as a function of total mass $M$ (top) and redshift $z$(bottom), for different source population models.}
    \label{fig:z}
\end{figure}

As shown in Eq.\ (\ref{tau}), the lensing probability increases with redshift, therefore one of the main properties that will determine the percentage of detectable strong-lensing events will be the redshift distribution of the populations. 

Fig.\ \ref{fig:z} shows the total mass and redshift distribution of all the populations considered here. Regarding the mass, we can see that the light-seed simulation (in orange) peaks in lower masses than the heavy-seed simulations, as expected. Regarding the redshift distribution, both light-seed and heavy-seed models have a wide redshift range up to $z=20$, with only the Q3-d (K+16) model showing preference to lower redshifts due to the delays in binary mergers. Importantly, the simulations from B+20 show higher rates for sources at redshift $z>15$, which suggests a higher rate of strong lensed events, as we confirm here.

\subsection{Lens Population}\label{sec:lenses}

We first model lensing caused by individual galaxies, which are the most common lenses. We will assume their radial mass profile to be axially symmetric. In particular, a common model is that of a singular isothermal sphere (SIS) density profile, which has been shown to provide a good approximation when estimating lensing probabilities \cite{Robertson:2020mfh}. In this case, the radial density profile is characterized by the velocity
dispersion, $\sigma$, as:
\begin{equation}
    \rho_{\rm SIS}(r)=\frac{\sigma^2}{2\pi G r^2}.
\end{equation}
All SIS lenses will always form two images, (+) and (-), each one with the following (de)magnifications
\begin{equation}\label{mags}
\mu_{\pm}= \left|1 \pm  \frac{1}{y}\right|,
\end{equation}
where $y=\theta_s/\theta_E$ is the dimensionless impact parameter, defined as the observed angular position of the source $\theta_{s}$ with respect to the optical axis (see Fig.\ \ref{fig:SLsetup}), and normalized by the angular Einstein radius $\theta_E$
\begin{equation}
    \theta_E=4\pi\left(\frac{ \sigma}{c}\right)^2\frac{D_{LS}}{D_S}.
\end{equation}

We see that if $y>1$ the two images are only slightly (de)-magnified which is the weak lensing regime, and if $y<1$ both images can be greatly magnified which is the strong lensing regime. Predicting other image multiplicities, such as four or five images, requires different lens mass profiles.

As we can see in Eq.\ (\ref{mags}), the magnifications depend on the location of the source, which is assumed to have a cumulative probability distribution\footnote{We quote the cumulative distribution since this is used to sample $y$ by performing a uniform inverse sampling.} of the form
\begin{equation}\label{impact}
    p(y)\propto y^2,
\end{equation} 
describing a uniform radial location in the 2D source plane. While the bulk of lensed events will correspond to those of weak lensing, since we are interested in strong lensing, we will focus on cases where $y\in [0,1]$.

We emphasize that different lens mass profiles could also be considered to improve the modeling of galaxy mass as function of radius. In particular, going beyond axially symmetric lenses can help include realistic effects of tidal fields and produce more than two images, as in the singular isothermal ellipsoid (SIE) lenses considered in previous similar studies \cite{Haris:2018vmn, Li:2018prc, Xu:2021bfn, Caliskan:2022wbh}. Other possibilities that can also produce more than two images are empirical mass profiles with shallower cores such as the Navarro-Frenk-White (NFW) model \cite{Navarro:1996gj}, as considered in some wave-optics lensing studies \cite{Savastano:2023spl, Brando:2024inp, Singh:2025uvp}. Nonetheless, as shown in \cite{Robertson:2020mfh}, an SIS assumption does provide reasonable lensing probabilities, although differences in parameter estimation of individual lensed GWs are expected to vary between different mass profiles. A more exhaustive exploration of different lens mass profiles and the production of more images will be left for future work.

In the case of SIS lenses, the time delay between the two images is given by \cite{1992grle.book.....S}
\begin{equation}\label{delays}
\Delta t = \Delta t_z \;y, \quad 
\Delta t_z \equiv \frac{32 \pi^2}{c} \left( \frac{\sigma_v}{c} \right)^4 \frac{D_L D_{LS}}{D_S} (1 + z_L),
\end{equation}
where $z_L$ is the redshift of the lens.

The main quantity characterizing the probability of strong lensing is the optical depth $\tau$, which is defined as the fraction of the celestial sphere that is gravitationally lensed, and interpreted as the probability of a given source to be strongly lensed. This optical depth only depends on the lens population, including the number density of lenses, and their cross sections. In general, $\tau$ can be calculated as \cite{1992grle.book.....S}:
\begin{equation}\label{Diff_tau}
    \frac{d^2\tau}{d z_L d \sigma} = \frac{dn (\sigma,z_L)}{d\sigma}\, S_{\rm cr}(\sigma,z_L,z_s) \frac{c dt }{d z_L}(1+z_L)^3,
\end{equation}
where $dn/d\sigma$ is the comoving number density distribution of SIS lenses (characterized solely by the velocity dispersion and lens redshift), $S_{\rm cr}$ is the lens cross section which in the strong lensing regime it is given by the area of the Einstein radius $S_{\rm cr}=\pi D_L^2(z_L) \theta_E^2(\sigma,z_L,z_s)$. In addition, we have assumed the lens distribution to be uniform in comoving volume leading to the scaling factors $(1+z_L)^3 (cdt/dz_L)$ with $cdt/dz_L=c/[H(z_L)(1+z_L)]$.

\subsubsection{Redshift-independent lenses}
Based on low-redshift observations of galaxy populations, it is customary to consider the galaxy population distribution to be independent of redshift \cite{2010MNRAS.404.2087B, Collett:2015roa, Ng:2017yiu, Haris:2018vmn, Li:2018prc, Yang_2021, Hannuksela:2019kle}. In that case, by matching the number density of galaxies and their (redshift-independent) velocity dispersions to the SDSS DR6 population of all types of galaxies\footnote{While early-type galaxies dominate strong lensing events due to their higher masses, late-type galaxies can make non-negligible contributions of up to $15\%$ of events \cite{2024SSRv..220...23L}, and thus a general approach should include all types of galaxies.}, \cite{2010MNRAS.404.2087B} found a velocity dispersion distribution following a modified Schechter function of the following form 
\begin{equation}\label{sigma}
    \frac{dn}{d\sigma}\equiv \phi(\sigma)=\phi_{*,0} \left(\frac{\sigma}{\sigma_{*,0}}\right)^\alpha \exp\left[-\left(\frac{\sigma}{\sigma_{*,0}}\right)^\beta \right]\frac{\beta}{\Gamma(\alpha/\beta)\sigma},
\end{equation}
with mean values $\phi_{*,0}=2.099\times 10^{-2}(h/0.7)^3 \text{Mpc}^{-3}$, $\sigma_{*,0}=113.78 \text{km/s}$, $\alpha=0.94$ and $\beta=1.85$\footnote{Note that different works have measured these parameters, obtaining somewhat different values depending on the data set and methodology. A commonly used set in previous GW lensing analyses comes from SDSS DR5 for early-type galaxies in \cite{Choi:2006qg} but here we use the subsequent study in \cite{2010MNRAS.404.2087B}.}.  %Due to the difficulty in modeling their velocity dispersions and mass distributions, their inclusion will be left for future work.

Eq.\ (\ref{sigma}) can be used to calculate the strong lensing optical depth of SIS lenses in Eq.\ (\ref{Diff_tau}) to obtain
%with amplification $\mu=\sqrt{\mu_+^2+\mu_-^2}>\mu_0$ \cite{Haris:2018vmn}\ml{cite others}
\begin{equation} \label{tau}
    \tau (z_s)=\int_{0}^{z_s}dz_L \int_{0}^{\infty}d\sigma \;\frac{d^2\tau}{dz_L d\sigma}=  F \left( \frac{D_c(z_s)}{cH_0^{-1}} \right)^3,%\left(\frac{2}{\mu_0}\right)^2
\end{equation}
where $F$ is a dimensionless normalization factor and $D_c(z_s)$ is the comoving distance from the source at redshift $z_s$. The value of $F$ is found to be $F=4.89\times 10^{-4}$ when using Eq.\ (\ref{sigma}). Notice that this value is slightly lower than that one used in previous GW studies \cite{Ng:2017yiu, Hannuksela:2019kle}, and hence we agree with the simulation-based analysis in \cite{Robertson:2020mfh} which found previously used values to be too high.
%has been taken to be between $(2-1.7)\times 10^{-3}$ in previous studies \cite{Ng:2017yiu, Hannuksela:2019kle}, but simulation studies of dark matter halos have shown these values to be overestimated \cite{Hilbert:2007jd, Robertson:2020mfh}. Here, we thus consider $F=\ml{3.77\times 10^{-3}}$ according to \cite{Robertson:2020mfh}.

In addition, the modeling of the lens population also encodes their redshift distributions. Following \cite{Haris:2018vmn}, for a given source at $z_s$, the lens redshift $z_L$ follows a distribution 
\begin{equation}\label{zL_dist}
    p(r)=30r^2(1-r)^2; \quad 0<r<1
\end{equation}
where $r= D_c(z_{L})/D_c(z_{s})$. This expression comes from using Eq.\ (\ref{sigma}) to obtain the optical depth per lensing redshift, $d \tau/d z_L=\int d\sigma (d^2\tau/dz_L d\sigma)$, which is found to follow the distribution in Eq.\ (\ref{zL_dist}) for SIS lenses. This distribution includes the effect of the lenses being distributed uniformly in comoving volume in addition to the lenses cross section distribution, which is known to be largest when the lens in neither close to the source nor to the observer. Indeed, the mode of the distribution (\ref{zL_dist}) is at $r=0.5$. Based in this, a high-redshift LISA source, say at $z_s=20$, would have a mostly likely lens at $z_L\approx 2$. Lenses are generally expected to lie at considerably lower redshifts than the sources but still lie beyond $z>1$.

\subsubsection{Redshift-dependent lenses}
Even though the previous model of redshift-independent velocity dispersion is justified by galaxy observations at redshift $z<1.5$, they are still applied to higher redshifts. Indeed, this method has been used to estimate the strong lensing rate of GW events for next-generation ground-based detectors \cite{Ng:2017yiu, Li:2018prc, Yang_2021, Xu:2021bfn} such as Einstein Telescope (ET) \cite{Abac:2025saz} and Cosmic Explorer (CE) \cite{Evans:2021gyd}, which are expected to observe events up to redshift $z_s\sim 10$. 

In the case of LISA, MBHBs can be observed up to $z_s\sim 20$ (see Fig.\ \ref{fig:z}) and thus it might be important to include redshift evolutions in the lens population modeling to perform realistic estimations. The previous LISA strong lensing rate analysis in \cite{Sereno:2010dr} considered both a redshift-independent and a redshift-dependent population model.

A common parametric redshift-dependent model of lens populations used in the literature \cite{Chae:2006kw, Oguri_2012, Geng:2021tiz, 2025MNRAS.537..779F} has a velocity dispersion function of the same form as Eq.\ (\ref{sigma}) but with evolving $\phi_*$ and $\sigma_*$ as:
\begin{equation}\label{sigma_z}
    \phi_* (z)= \phi_{*,0}(1+z)^{\nu_n}, \quad  \sigma_* (z)= \sigma_{*,0}(1+z)^{\nu_v},
\end{equation}
where $\nu_n$ and $\nu_v$ are two constant parameters characterizing the redshift evolution. From here on, this will be referred to as the $z$-dependent model 1.
%Previous analyses on galaxy strong lensing events have generally found support for $\nu_n\approx -1$ and $\nu_v\approx 0.25$ \ml{cite}, suggesting a decrease in number density and a mild increase in the velocity dispersions with redshift. 
In particular, by studying a sample of over 150 strong lensing events (caused by early-type galaxy lenses) with sources up to redshift 3.5, \cite{Geng:2021tiz} obtained median values
\begin{equation}\label{nus1}
     \nu_n=-1.18, \quad \nu_v=0.18.
\end{equation}
The LISA study in \cite{Sereno:2010dr} considered the redshift evolution in Eq.\ (\ref{sigma_z}) based on the simulation study in \cite{Chae:2006kw}, where the best-fit parameters where found to be $\nu_n=-0.229$ and $\nu_v=-0.01$, but here we will use the observation-based values in Eq.\ (\ref{nus1}). %Note that the values in Eq.\ (\ref{nus1}) imply that the number of galaxies decreases with redshift but their dispersion velocity increases

Nonetheless, since observations are limited in redshift, it is worth considering high-redshift lens models based on hydrodynamical numerical simulations.
In \cite{Torrey_2015} an analysis of dark matter halos in the Illustris simulation \cite{Nelson:2015dga} was performed to fit the cumulative velocity dispersion function up to $z=6$. From their fitted formulas one can obtain the velocity dispersion function $\phi_{hyd}(\sigma,z)$ and update the distribution $\phi(\sigma)$ in Eq.\ (\ref{sigma}) to include a redshift evolution as \cite{Oguri:2018muv}:
\begin{equation}\label{sigma_z2}
   \phi(\sigma,z) = \phi(\sigma) \frac{\phi_{hyd}(\sigma,z)}{\phi_{hyd}(\sigma,0)}, 
\end{equation}
where the expression for $\phi_{hyd}(\sigma,z)$ is provided in  Appendix \ref{app:VDF_z}. From here on, this model will be referred to as the $z$-dependent model 2. As discussed in \cite{Abe:2024rbf} this model may underestimate the number of galaxies at high redshifts, so here we consider it a conservative model. This is also confirmed in Appendix \ref{app:VDF_z}, where we show a comparison of velocity dispersion functions of the three models considered, and the $z$-dependent model 2 is shown to have orders of magnitude lower number densities at redshift $z=6$.

For simplicity, we will continue assuming that galaxies follow an SIS mass distributions at high redshift (even though this is not expected to be accurate). On the one hand, for the $z$-dependent model 1 in Eq.\ (\ref{sigma_z}), the lens redshift distribution function when the source is at $z_s$ is given by:
\begin{equation}\label{zL_dist1}
    p(z_L| z_s)\propto (1+z_L)^{\nu_n+4\nu_v}\frac{r^2(1-r)^2}{H(z_L)},
\end{equation}
where again $r=D_c(z_L)/D_c(z_s)$. This expression is equivalent to Eq.\ (\ref{zL_dist}) when $\nu_n=\nu_v=0$ and the sampled parameter is $r$ instead of $z_L$.

In addition, we find the optical depth to be given by:
\begin{align}\label{eq:tau2}
    \tau(z_s)&=16\pi^3 c \phi_{*,0} \frac{\Gamma\left(\frac{4+\alpha}{\beta}\right)} {\Gamma(\alpha/\beta)}\left(\frac{\sigma_{*,0}}{c}\right)^4  \nonumber\\
    &\times \int_0^{z_s} dz_L \; \frac{(1+z_L)^{2+\nu_n+4\nu_v}}{H(z_L)}\left(\frac{D_L D_{LS}}{D_S}\right)^2
\end{align}

On the other hand, for the $z$-dependent model 2 in Eq.\ (\ref{sigma_z2}), the lens redshift distribution function when the source is at $z_s$ is given by:
\begin{equation}\label{zL_dist2}
    p(z_L| z_s)\propto \frac{r^2(1-r)^2}{H(z_L)} \int_{\sigma_{\rm min}}^{\sigma_{\rm max}} d\sigma\; \sigma^4\phi(\sigma,z_L),
\end{equation}
which must be calculated numerically so we choose a relevant range of $(\sigma_{\rm min},\sigma_{\rm max})=(1,10^3)$km/s\footnote{We emphasize that the dependency of the results on this exact range of velocity dispersions is very weak, as long as the range covers most of the weight distribution of the function $\sigma^4\phi(\sigma,z_L)$.}. The optical depth is then given by:
\begin{equation}\label{eq:tau3}
    \tau(z_s)=\frac{16\pi^3}{c^{3}}\int_{0}^{z_s}dz_L \frac{(1+z_L)^{2}}{H(z_L)}\left(\frac{D_L D_{LS}}{D_S}\right)^2\int_{\sigma_{\rm min}}^{\sigma_{\rm max}} d\sigma\; \sigma^4\phi(\sigma,z_L).
\end{equation} 
We note that Eq.\ (\ref{eq:tau3}) is the general formula for the optical depth for SIS lenses. For the previous two models, integrations in $\sigma$ and/or $z$ were performed analytically and that is why the expressions look different in Eqs.\ (\ref{tau}) and (\ref{eq:tau2}).

\subsection{Lensing Comparisons}
\begin{figure}[h!]
\centering
\includegraphics[width=0.9\linewidth]{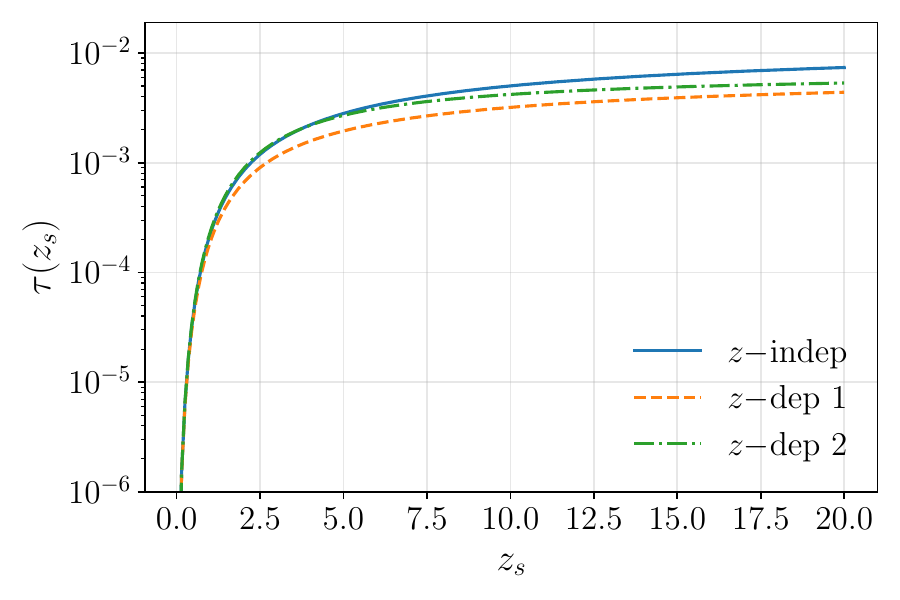}
\caption{Optical depth as function of source redshift, for the three lens population models: $z$-independent, $z$-dependent model 1, and  $z$-dependent model 2.}
\label{fig:taus}
\end{figure}

In this section, we perform a comparison between the three galaxy lens population models described in Sec.\ \ref{sec:lenses}. 
Fig.\ \ref{fig:taus} shows the optical depths obtained in the three models: $z$-independent (see Eq.\ (\ref{tau})), $z$-dependent 1 (see Eq.\ (\ref{eq:tau2})), and $z$-dependent 2 (see Eq.\ (\ref{eq:tau3})). 

From here we can see that, by construction, they are the same for low redshifts, but they differ at high redshifts. At $z=20$, the $z$-dependent model 1 gives the lowest optical depth of $4.39\times 10^{-3}$, whereas the $z$-dependent model 2 and the $z$-independent model give $5.33\times 10^{-3}$ and 
$7.38\times 10^{-3}$, respectively.

\begin{figure}[h!]
\centering
\includegraphics[width=0.9\linewidth]{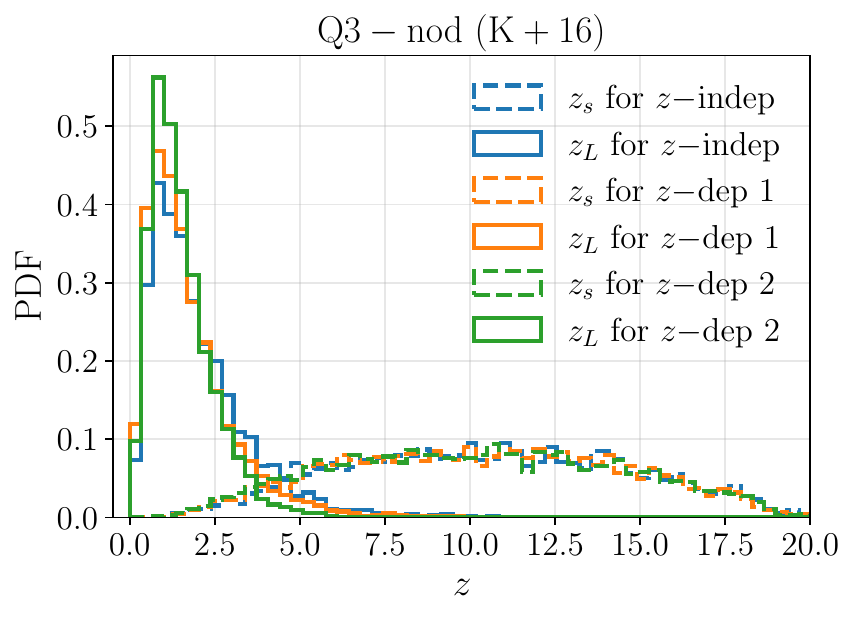}
\caption{Redshift distributions of lenses $z_L$ and lensed sources $z_s$ for the population Q3-nod (K+16).}
\label{fig:zlzs}
\end{figure}

In addition, Fig.\ \ref{fig:zlzs} shows the comparison on the lens redshift distributions in the three lens models (solid lines), together with the lensed source distribution (dashed lines). For concreteness, here we show the case when the lensed source population follows the Q3-nod (K+16) distribution, but other populations exhibit similar trends. Notice that the lensed source distribution shown here is obtained taking the intrinsic one in Fig.\ \ref{fig:z} and multiplied appropriately by the optical depth, for each one of the three lens models. The three models provided very similar lensed source distributions, but differences in the lens distributions are noticeable.

In particular, in the lens models with redshift evolution, we find the lenses to be at slightly lower redshifts, with the $z$-dependent model 2 being the one with less high-redshift lenses. In particular, in the $z$-independent lens model we find $26\%$ of the events to have $z_L<1$ and $81\%$ with $z_L<3$. In the $z$-dependent model 1, we find $33\%$ and $87\%$ of events with $z_L<1$ and $z_L<3$, respectively. In the $z$-dependent model 2, we find $34\%$ and $92\%$ of events with $z_L<1$ and $z_L<3$, respectively. Analogous lensing percentages for all source models are provided in Appendix \ref{app:zL_dist}.%\ml{values in appendix need to be updated}

We emphasize that these results describe the intrinsic population of lensed sources and lenses, not the observed ones. However, as we will see in Sec.\ \ref{sec:unlensed}, in the heavy-seed source models almost all events are detectable by LISA, and only in the light-seed source model the detection rate is low. In practice, this means that Fig.\ \ref{fig:zlzs} can be considered to also represent the detectable population of lensed sources and lenses for the Q3-nod (K+16) case. 

\section{Unlensed Event Rates}\label{sec:unlensed}
We start by first assessing the detectability of GW events without lensing effects. For this, we use the source population models discussed in Sec.\ \ref{sec:sources}, which provide distribution probabilities of the MBHBs in redshift, masses, and spins. Given these values of intrinsic parameters, we model GW signals using the \texttt{lisabeta} package \cite{Marsat:2020rtl, lisabeta2025}, which incorporates the time-delay interferometry (TDI) channels \cite{Tinto:2004wu} A, E and T, and the noise power spectrum follows the SciRDv1 model  \cite{lisa2018sciencereq}. We will also model the signal with the IMRPhenomHM waveform model which includes higher angular harmonics \cite{London:2017bcn}. This waveform describes quasi-circular non-precessing binary systems and needs an additional 6 extrinsic parameters to be specified: inclination angle, two angles for sky location, coalescence phase, polarization angle, and coalescence time. We randomize these extrinsic parameters and assess the detectability of GW events by calculating their mean signal-to-noise ratio (SNR) $\rho$ in a 4-year observation window. More details on this procedure are provided in Appendix \ref{app:unlensed}. We will set a detectability threshold of average SNR of $\rho>8$. 
%For calculating the SNR we use \texttt{LISAanalisistools} \cite{michael_katz_2024_10930980}, which incorporates the time-delay interferometry channels A, E and T, and the noise power spectrum follows the SciRDv1 model \ml{cite}. 

The intrinsic astrophysical and detectable number of events in 4 years are quoted in Table \ref{tab:Ndetectedevents}. These estimations are in agreement with previous studies, e.g.\ \cite{Barausse:2023yrx}. From these results we see that light-seed models tend to predict a lower fraction of detectable events than heavy-seed models, due to their lower mass distributions \cite{Sadiq:2024xsz}. From Table \ref{tab:Ndetectedevents} we obtain for heavy-seed models a detection fraction always over 96\%, while in the light seed model PopIII-d (K+16) it is about 23\%.

\begin{table}[h!]
%\centering
\begin{tabular}{|l|c|c|}
\hline
\textbf{Source Model} &  \textbf{Intrinsic} & \textbf{Detected}\\
\hline
\hline
PopIII-d (K+16)            & 1,410     & 321     \\ \hline   % 22.77\%
Q3-d (K+16)                & 74       & 73     \\ \hline   % 98.65\%
Q3-nod (K+16)              & 657      & 650   \\ \hline   % 98.93\%
HS-nod-noSN (B+20)         & 39,364    & 37,979    \\ \hline  % 96.48\%
HS-nod-SN (B+20)           & 36,790    & 35,510    \\ \hline   % 96.52\%
HS-nod-SN-high-accr (B+20) & 9,407     & 9,075   \\   %9 6.47\%
\hline
\end{tabular}
\caption{Average number of intrinsic and unlensed detected events, for each simulation of population mergers, in a 4-year LISA mission. }
\label{tab:Ndetectedevents}
\end{table}

\begin{figure}[h!]
    \centering
        \includegraphics[width=1\linewidth]{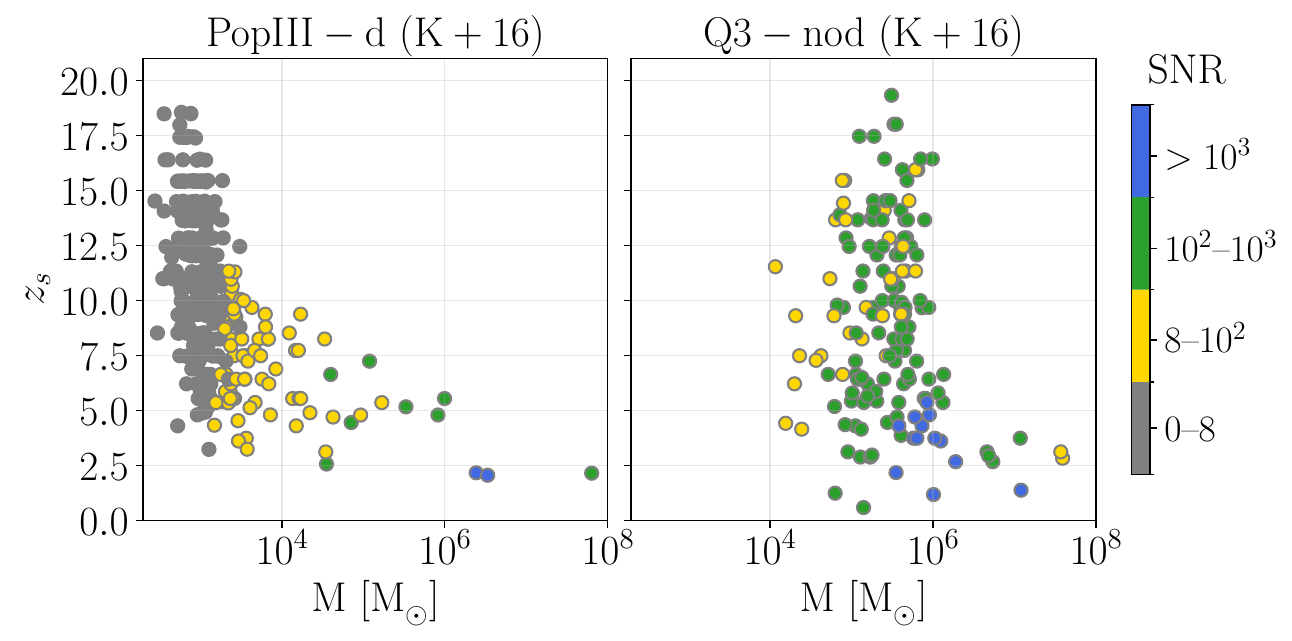}
    \caption{Intrinsic events for PopIII-d (K+16) and Q3-nod (K+16) simulations in 1 year, as function of total mass $M$ and source redshift $z_s$. Colors indicate the expected mean SNR, with grey dots showing the undetectable events.}
    \label{fig:snr_unlensed}
\end{figure}

%The astrophysical and unlensed detectable event rates in 4 years, in redshift and total mass are shown in Fig.\ \ref{fig:popIII_detect}\ml{to be updated}, where it is confirmed that the low-mass events are lost.
In Fig.\ \ref{fig:snr_unlensed} we show the mean SNR distribution of the intrinsic population of events in 1 year of LISA observing time, for the simulations PopIII-d (K+16) and Q3-nod (K+16) as examples. Here we see that most light-seed events become undetectable, while 90\% of the detectable events have $8<\rho<100$. In the Q3-nod (K+16) model we see that most events are detectable with 24\% of them having  $8<\rho<100$, and 76\% of them with $\rho>10^2$ due to the higher masses. In the other heavy-seed source models we find the fraction of detectable events with $8<\rho<100$ to be: 3.9\% for Q3-d (K+16), 58\% for HS-nod-noSN (B+20), 58\% for HS-nod-SN (B+20), and  63\% for  HS-nod-SN-high-accr (B+20). %\ml{discuss whether it is better to show $8<\rho<10^3$}.

\section{Lensed Event Rates}\label{sec:rates}

\subsection{Event Number}\label{sec:numbers}
In this section, we show the results on the number of strong-lensed events. In order to calculate this, we take the astrophysical population distribution and weigh it appropriately, as a function of redshift, by the lensing optical depths $\tau(z_S)$ shown in Fig.\ \ref{fig:taus}. The number of intrinsic strongly lensed events in a 4-year LISA mission are provided in Table \ref{tab:SL_intrinsic}.
\begin{table}[h!]
\centering
\begin{tabular}{|l|c|c|c|c|}
\hline
\textbf{Source Model} &   \textbf{$z$-indep}& \textbf{$z$-dep 1}  & \textbf{$z$-dep 2} \\
\hline
\hline
PopIII-d (K+16)            & 6.9    & 4.4     &  5.7     \\\hline
Q3-d   (K+16)              & 0.20    & 0.14     & 0.19      \\\hline
Q3-nod   (K+16)            & 2.9    & 1.9     & 2.4      \\\hline  
HS-nod-noSN (B+20)         & 231  & 143   & 181    \\ \hline
HS-nod-SN (B+20)           & 216  & 133   & 169    \\ \hline
HS-nod-SN-high-accr (B+20) & 51   & 32    & 41     \\ \hline
\end{tabular}
\caption{Number of intrinsic strongly lensed events in a 4-year LISA mission.}
\label{tab:SL_intrinsic}
\end{table}

Here we see that the number of lensed events in 4 years can have a wide range: between 0.14 (approximately one every 30 years) and 231 (approximately one every week). As expected, the $z$-independent lens model generally predicts more lensed events than the two $z$-dependent lens models, due to its higher number of high-redshift lenses. Then, the $z$-dependent model 2 predicts more lensed events than the $z$-dependent model 1, due to the former having slightly higher optical depth, as shown in Fig.\ \ref{fig:taus}. Nonetheless, the predictions of the three lens models are comparable, with fractional differences on the event rates of up to $38\%$. 
%fractional differences: (6,9-4,4)/6,9=36% for PopIII; 32% for Q3d; 34% for Q3nod; 38% for HSnodnoSN; 38% for HSnodSN; 37% for HSnodSNhighaccr.
%
Since the optical depth is around $\tau \sim 10^{-3}-10^{-2}$ for $z_s\gtrsim 1$, the rates obtained in Table \ref{tab:SL_intrinsic} are about a factor $10^{-3}-10^{-2}$ lower than the intrinsic unlensed event rates presented in Table \ref{tab:Ndetectedevents}.

From the intrinsically lensed population distributions, we mock the detectability of the two strong-lensed images in 4 years. In particular, we take 5,000 events with intrinsic parameters $(m_1,m_2,z,a_1,a_2)$, and for each one of them we perform 100 realizations of lensing by sampling the impact parameter from Eq.\ (\ref{impact}). In each realization, we calculate the  SNR associated to each one of the strong-lensed images as:
\begin{equation}
    \rho_+=\sqrt{\mu_+(y)}\; \rho; \quad  \rho_-=\sqrt{\mu_{-}(y)}\; \rho
\end{equation}
where $\rho$ is the unlensed mean SNR (as discussed in Sec.\ \ref{sec:unlensed}), and the magnifications are in Eq.\ (\ref{mags}). In each of the 100 realizations we calculate the number of events with: only 1 detectable image, 2 detectable images, and events that went above threshold due to strong lensing and would be otherwise undetectable (i.e.\ boosted). 

In Table \ref{tab:SLrates} we show the mean values across the 100 realizations for the three number of events obtained, after rescaling the 5,000 sampled lensed events to the appropriate intrinsic lensed events in 4 years (see Table \ref{tab:SL_intrinsic}). As a consequence, the numbers quoted in Table \ref{tab:SLrates} provide the mean expectation of the number of detectable strong-lensed events in 4 years, for the different source and lens population scenarios considered in this paper. From Table \ref{tab:SLrates} the total number of lensed events can be calculated as the sum of the second and third columns.

\begin{table}[h!]
\centering
\begin{subtable}{\linewidth}
\begin{tabular}{|l|c|c|c|}
\hline
\multicolumn{4}{|c|}{\textbf{$z$-independent lens model}}\\
\hline
\textbf{Source Model} &   \textbf{1 Detected}& \textbf{2 Detected}  & \textbf{Boosted} \\
\hline
\hline
PopIII-d (K+16)            & 1.9   & 0.95   & 1.7  \\ \hline
Q3-d   (K+16)              & $1.1\times 10^{-3}$   & 0.19   & $2.4\times 10^{-4}$  \\ \hline
Q3-nod   (K+16)            &  0.084  & 2.8   & 0.013  \\ \hline  
HS-nod-noSN (B+20)         & 21  & 210 & 2.5  \\ \hline
HS-nod-SN (B+20)           & 18  & 196 & 1.8  \\ \hline
HS-nod-SN-high-accr (B+20) & 4.7   & 46  & 0.41  \\ \hline
\end{tabular}
\caption{Lensed events for $z$-independent lens model.}
\label{tab:SLrates_zind}
\end{subtable}
%%%%%%%%%%%%%%%%
\begin{subtable}{\linewidth}
\begin{tabular}{|l|c|c|c|}
\hline
\multicolumn{4}{|c|}{\textbf{$z$-dependent lens model 1}}\\
\hline
\textbf{Source Model} &   \textbf{1 Detected}& \textbf{2 Detected}  & \textbf{Boosted} \\
\hline
\hline
PopIII-d (K+16)            & 1.2  & 0.62   & 1.0   \\ \hline
Q3-d   (K+16)              & $6.9\times 10^{-4}$  & 0.13   & $1.4\times 10^{-4}$   \\ \hline
Q3-nod   (K+16)            & 0.052  & 1.8   & $3.7\times 10^{-3}$  \\ \hline  
HS-nod-noSN (B+20)         & 13 & 129 & 1.5   \\ \hline
HS-nod-SN (B+20)           & 11 & 121 & 1.1   \\ \hline
HS-nod-SN-high-accr (B+20) & 2.9  & 29  & 0.30   \\ \hline
\end{tabular}
\caption{Lensed events for $z$-dependent lens model 1.}
\label{tab:SLrates_zdep1}
\end{subtable}
%%%%%%%%%%%%%%%%%%
\begin{subtable}{\linewidth}
\begin{tabular}{|l|c|c|c|}
\hline
\multicolumn{4}{|c|}{\textbf{$z$-dependent lens model 2}}\\
\hline
\textbf{Source Model} &   \textbf{1 Detected}& \textbf{2 Detected}  & \textbf{Boosted} \\
\hline
\hline
PopIII-d (K+16)            & 1.6  & 0.80    & 1.4  \\ \hline
Q3-d   (K+16)              & $1.0\times 10^{-3}$  & 0.18    & $2.1\times 10^{-4}$  \\ \hline
Q3-nod   (K+16)            & 0.071  & 2.3    & $8.2\times 10^{-3}$ \\ \hline  
HS-nod-noSN (B+20)         & 16 & 164  & 1.6  \\ \hline
HS-nod-SN (B+20)           & 14 & 154  & 1.5  \\ \hline
HS-nod-SN-high-accr (B+20) & 3.8  & 37   & 0.39  \\ \hline
\end{tabular}
\caption{Lensed events for $z$-dependent lens model 2.}
\label{tab:SLrates_zdep2}
\end{subtable}
\caption{Mean number of strongly lensed events: with only 1 detectable image (2nd column), with 2 detectable images (3rd column), and those that went above threshold due to lensing or ``boosted'' (4th column), in a 4-year  LISA mission.}
\label{tab:SLrates}
\end{table}

From Table \ref{tab:SLrates} we see that the three different lens models provide comparable results, with fractional differences in the predicted event rates of up to nearly $40\%$. When comparing Table \ref{tab:SLrates}  with Table \ref{tab:SL_intrinsic} we find that in the Pop-III (K+16) source model many intrinsically lensed events are not detectable, with about $40\%$ of them being detectable, whereas for the heavy-seed source models roughly $99\% $ of the intrinsically lensed events are detectable. 

%\ml{mention the $\%$ of detectable lensed events, compared to the intrinsic number of lenses events, and compared to the number of unlensed events}
The rates quoted in Table \ref{tab:SLrates} range between $0.13-231$ in 4 years, and this upper bound is much higher compared to the previous result obtained in \cite{Sereno:2010dr}, where up to 4 strong-lensed events in a 5-year LISA mission where predicted. This difference mainly lies on the wider range of source models considered in this paper. The event numbers obtained here correspond to $0.2\%-0.9\%$ of all unlensed detectable events (see Table \ref{tab:Ndetectedevents}), depending on the source and lens model. Even though in these tables we only quote the mean, we have confirmed that the dispersion across the 100 realizations of these event rates is at least one order of magnitude smaller than the mean. 

In Table \ref{tab:SLrates} we also present the expected mean number of boosted events. In the Pop-III (K+16) source model, among all the detectable lensed events, the fraction of boosted events can be as high as $60\%$, whereas in the heavy-seed source models the boosted events represent a small fraction of detectable lensed events, reaching up to $1\%$. These results are expected due to the different mass distributions between light-seed and heavy-seed models. %Note that in the $5\times 10^{5}$ simulated lensed events performed here, no boosted event was found in the Q3-d (K+16) source model, indicating their fraction to be lower than $10^{-5}$ and hence effectively negligible.

\begin{figure}[h!]
    \centering
        \includegraphics[width=1\linewidth]{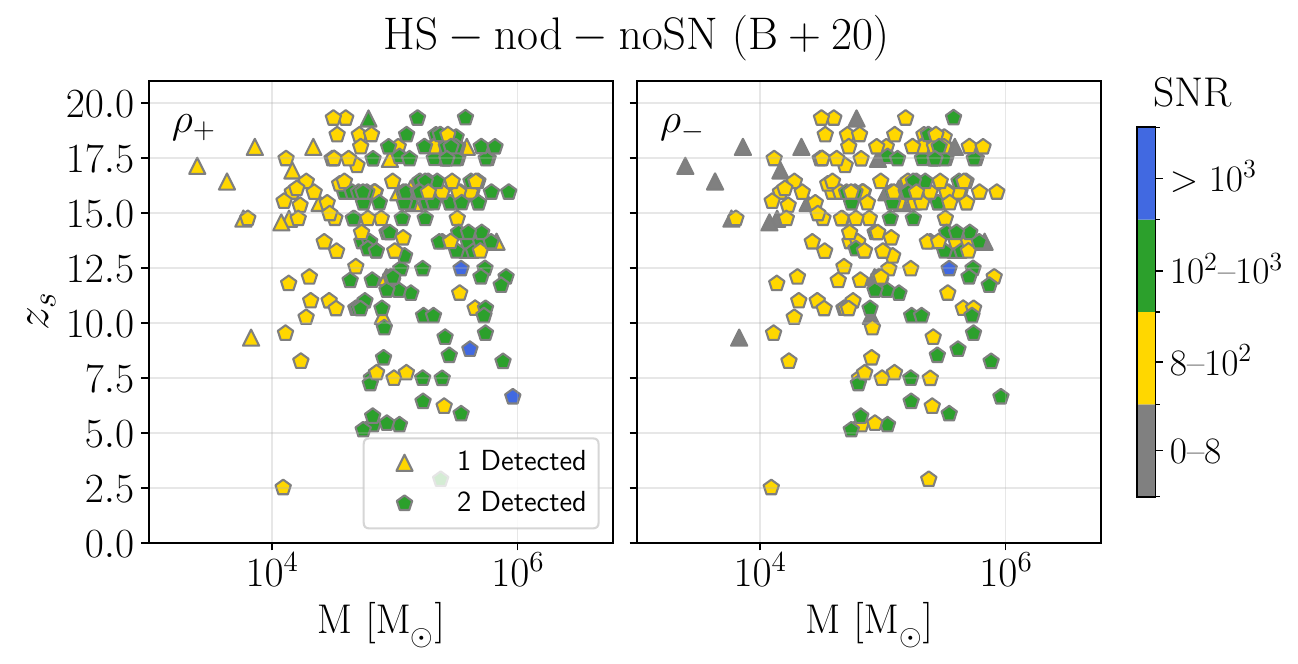}
    \caption{Lensed events for the HS-nod-noSN (B+20) source model, in 4 years, for the $z$-independent lens model, and their expected strong-lensed mean SNR $\rho_+$ (on the left) and $\rho_-$ (on the right). The events are shown as function of their redshift and total mass.}
    \label{fig:snr_lensed}
\end{figure}

Finally, in Fig.\ \ref{fig:snr_lensed} we show the mean SNR of the two images, for the HS-nod-noSN (B+20) source model in 4 years, in the $z$-independent lens model. We illustrate this case because it yields the highest overall number of lensed events. In pentagons we see the SNR of the events that have two detectable images, and in triangles we show those with only one detectable image (the (+) image). The color scheme describes the expected mean SNR, with gray showing undetectable images.

We see that due to strong lensing, the first image (+) always has higher SNR than the second image (-). In particular, for the case shown in Fig.\ \ref{fig:snr_lensed}, 27\% of the first images have $\rho>10^2$, compared to the 60\%  of second images. Notice that in the heavy-seed source models the two images of lensed events are typically detectable, with at most 10\% of cases having a missed second image. On the contrary, the light-seed source model predicts nearly two times more observed lensed events with only the first image detectable compared to the case with both images. Here, we find that about 40\% of the missed second images may still have SNR between 5 and 8. Therefore, mitigation strategies to solve this problem can be taken since follow-up analyses of sub-threshold events can be performed \cite{LIGOScientific:2021izm, LIGOScientific:2023bwz, Janquart:2023mvf}. In addition, the prediction that second images lead to a $\pi/2$ phase shift and, as a consequence, possible distortions when higher-angular modes are relevant \cite{Ezquiaga:2020gdt} in the GW waveform (as is the case for LISA) will also help find second images.

\subsection{Time Delays}

Previous studies \cite{Haris:2018vmn, Li:2018prc, Yu:2020agu, LIGOScientific:2021izm, LIGOScientific:2023bwz} for ground-based GW detectors have shown that information on time-delay distributions of images can be used to confidently identify strong-lensed GW events, and distinguish them from apparently similar events from different sources. For this reason, here we include an analysis on time delays of the strong-lensed events expected in LISA. 

We consider the same 5,000 intrinsic lensed events with parameters $(m_1,m_2,z_S,a_1,a_2)$ used in the previous subsection \ref{sec:numbers}. For each event, we sample both the redshift and velocity dispersion of the lens and calculate $\Delta t$ in Eq.\ (\ref{delays}), to produce a time delay probability distribution of intrinsically lensed events. 
For the lens redshift $z_L$, we draw samples from the redshift distributions in Eqs.\ (\ref{zL_dist}), (\ref{zL_dist1}) and (\ref{zL_dist2}) described in Sec.\ \ref{sec:lenses}. For the lens velocity dispersion, we draw samples from $\sigma^4\phi(\sigma,z_L)$, using Eqs.\ (\ref{sigma}), (\ref{sigma_z}) and (\ref{sigma_z2}).

\begin{figure}[h!]
    \centering
    \begin{subfigure}{0.49\textwidth}
        \centering
        \includegraphics[width=\linewidth]{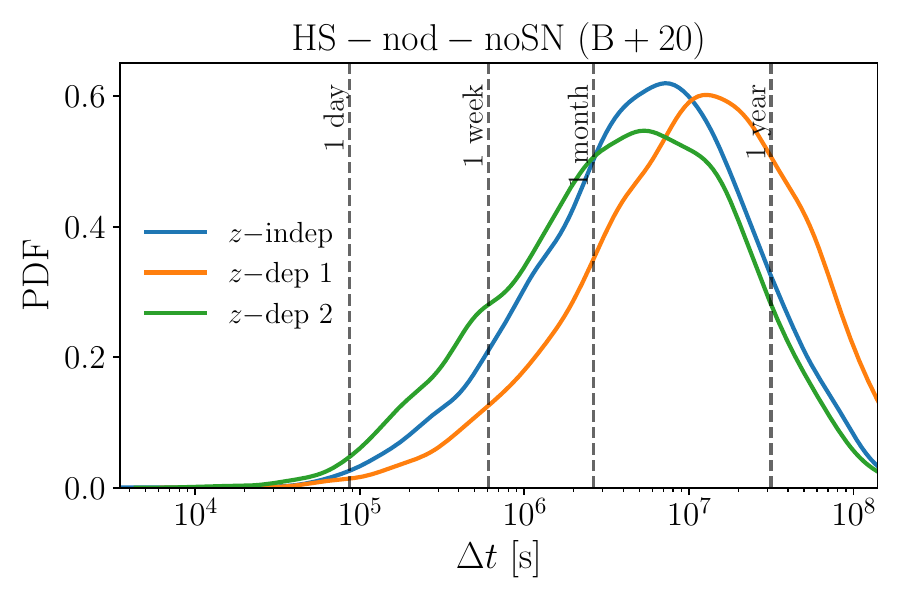}
    \end{subfigure}
    \caption{Probability distribution of the time delays between the two lensed images, for the HS-nod-noSN (B+20) source model, in the three lens scenarios with their KDE fits. }
    \label{fig:td}
\end{figure}

Fig.\ \ref{fig:td} shows the time delay distribution for the example of HS-nod-noSN (B+20) source population, for the three lens models. 
As observed, the peak is of a few months, but the three lens models can predict different tails for very short ($<1$ week) or very long ($>1$ yr) time delays. For this source model, in the range of 1 week $<\Delta t<$ 1 year, we find 81\% of events in the $z$-independent lens model, 74\%  in the $z$-dependent 1 model, and 78\% in the $z$-dependent 2 model.

\begin{figure}[h!]
    \centering
    \begin{subfigure}{0.49\textwidth}
        \centering
        \includegraphics[width=\linewidth]{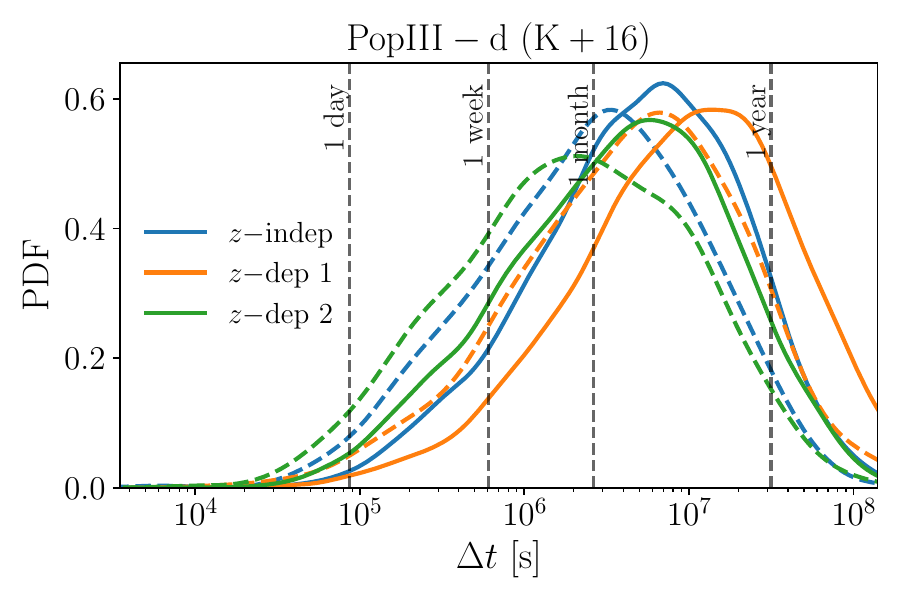}
    \end{subfigure}
    \caption{Probability distribution of the time delays between the two images of the intrinsic (solid lines) vs detectable (dashed lines) lensed events, for the PopIII-d (K+16) source model, in the three lens scenarios with their KDE fits.}
    \label{fig:td2}
\end{figure}

Other heavy-seed models provide qualitatively similar results as Fig.\ \ref{fig:td}. We recall that this result describes the time delay of the intrinsically lensed events but, as discussed previously, heavy-seed models predict about 99\% of lensed events to be detectable so the result shown here translates directly to the detectable events. This is not the case of the light-seed PopIII-d (K+16) model, where only 40\% of lensed events are detectable. For this reason, in Fig.\  \ref{fig:td2} we compare the time delay distribution of the intrinsic (solid lines) versus detectable (dashed lines) lensed events. In this case, we only considered lensed events where both images are detectable.

From here, we can see that the intrinsic time delay distribution of this light-seed model is qualitatively similar to that of the heavy-seed model in Fig.\ \ref{fig:td}, with peaks of a few months. Nonetheless, the detectable distributions have support for smaller time delays due to the fact that detectable sources have lower redshift, and thus produce lower time delays. The peaks of the intrinsic versus detected distributions shift by $3.5\times 10^6 $ s (or 5.8 weeks) in the $z$-independent lens model, $6.9\times 10^6 $ s (or 11.4 weeks) in the $z$-dependent 1 model, and $3.6\times 10^6 $ s (or 6 weeks) in the $z$-dependent 2 model.

%Despite being different scenarios, the three models show qualitatively similar distributions, concentrated in a range of $\sim10^6$ to $\sim10^{7.5}$ seconds, i.e., from about couple weeks to one year. The only notable difference is that the Q3d simulation tends to produce slightly shorter time delays. This is consistent with the fact that the sources in this model are located at lower redshifts compared to the other populations. This prediction is encouraging from an observational point of view, since it implies that, in most cases, both signals could potentially be detected within the nominal 4 years duration of the LISA mission. 

%\ml{Need to compare results with previous studies}

\section{Discussion}\label{sec:discussion}
In this work, we have conducted a comprehensive analysis of the expected strong gravitational lensing rates for massive black hole binary (MBHB) mergers detectable by LISA over a 4-year mission duration. The detection of strong-lensed events could allow for improved parameter estimations to perform tests of gravity and dark matter, as well as improving sky localization and thus opening the possibility for multi-messenger studies with EM follow-ups. The results obtained here provide a significant expansion of previous estimates. We incorporate the latest MBHB population models and consider the unique challenges posed by LISA's high-redshift detection capabilities extending to redshift $z \sim 20$. In particular, we analyzed six different MBHB source models in combination with three different high-$z$ galaxy lens models. 

We find that the expected number of detectable strong-lensed MBHB events ranges from 0.13 to 231 events over 4 years, depending on the source population model and lens population assumptions. This wide range reflects the substantial uncertainties in both the formation channels and evolutions of MBHBs. 
These predicted lensed event rates correspond to 0.2\% to 0.9\% of all unlensed detectable MBHB mergers, establishing strong lensing as a rare but scientifically valuable phenomenon for LISA observations. 

The signal-to-noise ratio (SNR) of these lensed events can be as high as $10^4$, with the light-seed source population model producing generally lower SNRs and thus predicting the miss of second lensed images, whereas heavy-seed source population models generally predict higher SNRs and hence two lensed images being detectable most of the time. Due to this difference in the SNRs for the light and heavy seeds, we also find that up to 60\% of detectable lensed events in light-seed scenario (and about 1\% in heavy-seed scenarios) would be ``boosted" above the detection threshold solely due to lensing magnification, representing events that would otherwise remain undetectable.

In addition, our analysis of three galaxy lens population models--one redshift-independent and two redshift-dependent evolutionary scenarios--reveals that lens evolution introduces variations of approximately 40\% in the predicted event rates. The most common scenario considered previously in the literature is the redshift-independent one, which generally predicts the highest rates due to its assumption of constant galaxy number densities at all redshifts.

In addition, we performed an analysis on the expected time delays between two lensed images, as this may be used for future detectability studies to improve confidence in the lensing hypothesis between a pair of events. Across different heavy-seed source and lens models, we find between 72\%-81\% of lensed events to have a time delay between 1 week and 1 yr, peaking typically at several months. In the light-seed source population model, the time delays are lower since their lower masses cause these sources to be at lower redshift in order to be detectable.

While the largest uncertainties in the type of analysis performed in this paper will remain to be the MBHBs populations, we have made several simplifying assumptions that warrant future investigation. The adoption of singular isothermal sphere lens models, while computationally tractable, neglects the complexity of real galaxy mass distributions. More sophisticated lens models, such as singular isothermal ellipsoids, would produce four-image configurations and allow us to improve our rate predictions.

Additionally, our detectability assessment relies on sky-averaged SNRs without fully accounting for the temporal separation and different sky locations of multiple images. Detailed simulations incorporating these effects, along with realistic noise realizations and data analysis procedures, would refine our predictions.

The substantial uncertainties in both source and lens populations highlighted in this work underscore the importance of continued theoretical and observational efforts to better understand MBHB formation channels and galaxy evolution at high redshift. Future gravitational wave observations, particularly from Pulsar Timing Arrays and eventually LISA itself, will shed further light on these uncertainties.

Various plots for the population models not shown explicitly in this paper can be found in \cite{Github_JG_2025}. 

\section*{ACKNOWLEDGMENTS}
We thank Timo Anguita and Alberto Mangiagli for useful discussions. JG and ML were supported by Fondecyt Iniciación 11250105 grant. 

\appendix

\section{Unlensed Detectable Events}\label{app:unlensed}

This section describes our procedure for calculating the mean unlensed signal-to-noise ratio (SNR) of gravitational wave events detectable by LISA. Our analysis uses the IMRPhenomHM waveform model, which describes non-precessing binary black holes with aligned or anti-aligned spins and includes higher-order angular modes.

Each gravitational wave event is characterized by 11 parameters. On the one hand, we have 5 intrinsic parameters $(m_1,m_2,z,a_1,a_2)$ describing the masses, redshift, and spins of the BHs, which are drawn from the semi-analytical source models described in Section \ref{sec:sources}. On the other hand, we have 6 extrinsic parameters describing the observation geometry and timing, which are sampled as follows:
 \setlist{nolistsep}
\begin{itemize}[noitemsep]
    \item Coalescence time $t_c$: sampled uniformly between $t_c\in [0,4]$ yrs.
    \item Coalescence phase $\phi_c$: sampled uniformly between $\phi_c\in [0,2\pi]$.
    \item Inclination angle $\iota$ and polarization angle $\psi$: sampled isotropically in a sphere with $\psi\in [0,2\pi]$ and $\iota\in [0,\pi]$.
    \item Sky localization angles, ecliptic longitude $\lambda$ and latitude $\beta$: sampled isotropically in a sphere with $\lambda\in [0,2 \pi]$ and $\beta\in [-\pi/2,+\pi/2]$.
\end{itemize}
We initially draw a sample of 10,000 intrinsic parameters, from the raw distributions for the different source models (see Fig.\ \ref{fig:z}). For each set of intrinsic parameters, we generate $N=100$ random realizations of the extrinsic parameters and calculate the corresponding optimal SNR values. Our SNR calculations make two simplifying assumptions: we neglect signal confusion from overlapping sources and assume noiseless data. We compute the total SNR using the \texttt{lisabeta} package, which combines contributions from all three LISA TDI channels (A, E, and T).

\begin{figure}[h!]
\centering
\includegraphics[width=0.9\linewidth]{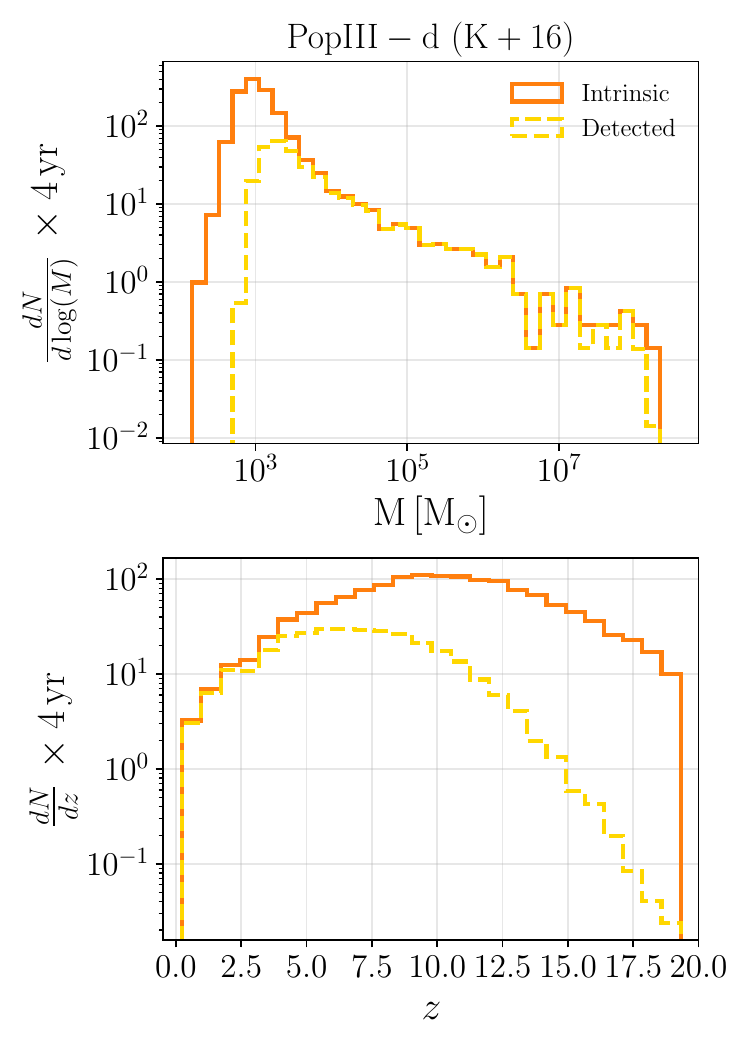}
\caption{Comparison of intrinsic vs unlensed detected distribution in redshift and total mass of the PopIII-d (K+16) source model, in a 4 year duration mission.}
\label{fig:popIII_unlensed}
\end{figure}

From the $N$ SNR calculations for each intrinsic parameter set $(m_1,m_2,z,a_1,a_2)$, we count  $N_{det}$ cases that exceed our detectability threshold of $\rho>8$. This gives us a detection probability  $N_{det}/N$ as a function of the intrinsic parameters, which we use to reweight the raw distributions from the population simulations. Figure \ref{fig:popIII_unlensed} illustrates the resulting detectable event distribution over 4 years for the PopIII-d (K+16) simulation, as an example. In the heavy-seed models, both intrinsic and detected distributions are very similar, given that almost all events are detected (see Table \ref{tab:Ndetectedevents}).

We integrate the detection-weighted distributions to obtain the total number of detectable events over 4 years, as summarized in Table \ref{tab:Ndetectedevents}. 

Additionally, we calculate the mean SNR for each intrinsic parameter combination by averaging over all 
$N$ extrinsic parameter realizations, which is used in Fig.\ \ref{fig:snr_unlensed}. Note that one may also estimate the number of detectable events by setting a detectability threshold of 8 on the \textit{mean} SNR. This would provide slightly different detection rates to the ones reported in Table \ref{tab:Ndetectedevents}, with less than $10\%$ differences.
%popIII 293.427; Q3d: 72.942; Q3nod: 652.817; HSnodnoSN: 38793.030; HSnodSN: 36270.958; HSnodSNhighaccr: 9285.550

\section{Lens Population}

\subsection{Velocity Dispersion Function}\label{app:VDF_z}
In this section, we describe in detail the lens population properties in $z$-dependent model 2.

\begin{figure}[h!]
    \centering
        \centering
  \includegraphics[width=0.98\linewidth]{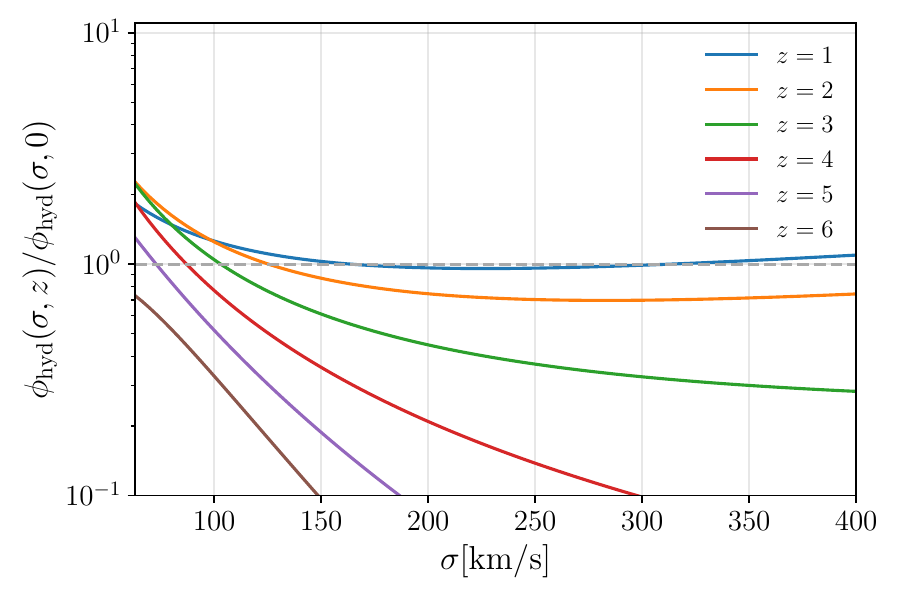} 
    \caption{Fractional evolution of the velocity dispersion function in Eq.\ (\ref{sigma_z2}), at several redshifts, as a function of the velocity dispersion $\sigma$.}
    \label{fig:vdf_hydro}
\end{figure}

The cumulative dispersion function in \cite{Torrey_2015} is fitted in log base-10 space as\footnote{Notice that \cite{Torrey_2015} has typos and thus we follow their github publication in \url{https://github.com/ptorrey/torrey_cmf}.}
\begin{equation}\label{CVDF_torrey}
 \log n= a + \alpha\log (\sigma/\sigma_*) +\beta\log^2 (\sigma/\sigma_*)-e^{\log (\sigma/\sigma_*)},
 % A \left(\frac{\sigma}{\sigma_*}\right)^{\alpha +\beta \log(\sigma/\sigma_*)} e^{-\sigma/\sigma_*} , 
\end{equation}
where $n$ is in units of $\text{Mpc}^{-3}$, and the galaxies studied had $\sigma> 63$ km/s. The parameters $a$, $\alpha$, $\beta$, and $\sigma_*$ are allowed to evolve quadratically in redshift as:
\begin{align}\label{sigma_coeffs}
    & a= a_{0}+ a_{1}\,z + a_{2}\,z^2 \; , \nonumber\\
    & \alpha= \alpha_{0}+ \alpha_{1}\,z + \alpha_{2}\,z^2, \nonumber\\
    & \beta= \beta_{0}+ \beta_{1}\,z + \beta_{2}\,z^2, \nonumber\\
    &\sigma_* = 10^{\gamma_{0}+ \gamma_{1}\,z + \gamma_{2}\,z^2} \; \text{km/s},
\end{align}
The best-fit values obtained in \cite{Torrey_2015} are
\begin{align}
  & a_0=7.391498, \; a_1=5.729400,\; a_2=-1.120552, \nonumber \\
    & \alpha_0=-6.86339338, \; \alpha_1=-5.27327109, \; \alpha_2=1.10411386,\nonumber  \\
    & \beta_0=2.85208259, \; \beta_1=1.25569600, \; \beta_2=-0.28663846,\nonumber \\
     & \gamma_0=0.06703215, \; \gamma_1=-0.04868317, \; \gamma_2=0.00764841.
\end{align}
Since for strong lensing estimates we need the velocity dispersion function distribution, we take the derivative of $n$ in Eq.\ (\ref{CVDF_torrey}) to obtain
\begin{align}
    \phi_{\rm hyd}(\sigma,z)=-\frac{dn}{d\sigma}=-\frac{n}{\sigma} \left( \alpha+ 2\beta\log(\sigma/\sigma_*)-e^{\log(\sigma/\sigma_*)}\right),
\end{align}
where $n$ is the expression in Eq.\ (\ref{CVDF_torrey}). This is the function that will be used in Eq.\ (\ref{sigma_z2}). The factor that will determine the redshift-evolution correction of the VDF will be given by $ \phi_{\rm hyd}(\sigma,z)/ \phi_{\rm hyd}(\sigma,0)$, which is shown in Fig.\ \ref{fig:vdf_hydro}. 
As expected from hierarchical galaxy evolution, at high redshift there are more galaxies with low velocity dispersions than large velocity dispersions. The normalization at $z=0$ has been included to make this model match the SDSS catalog number density as in Eq.\ (\ref{sigma_z2}), and to have a self-consistent comparison between the three lens models.

\begin{figure}[h!]
    \centering
        \centering
        \includegraphics[width=0.98\linewidth]{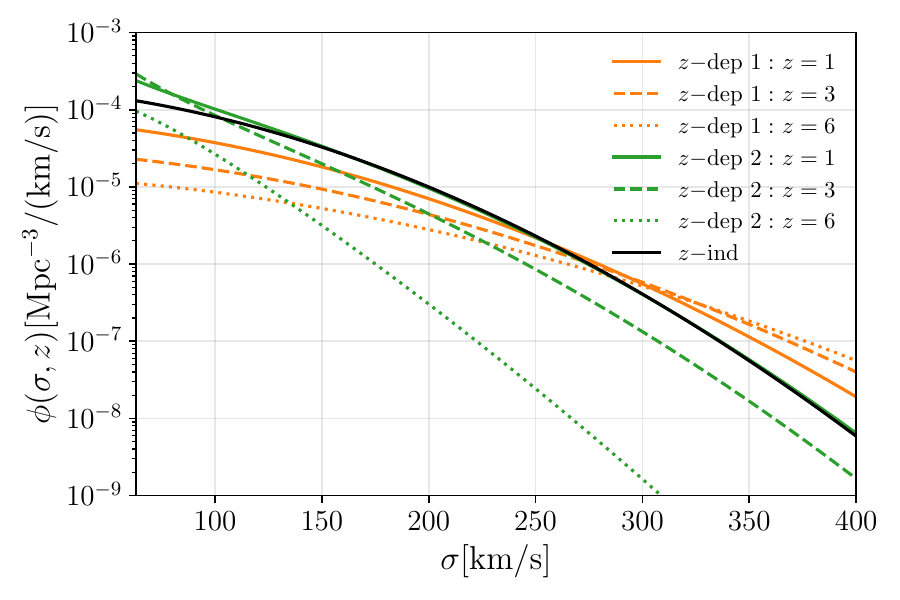} 
    \caption{Comparison of velocity dispersion functions at different redshifts, as a function of the velocity dispersion $\sigma$. At $z=0$ the three models are the same, by construction.}
    \label{fig:vdf_comp}
\end{figure}

A comparison of the three VDF models: $z$-independent, $z$-dependent model 1 and $z$-dependent model 2 can be found in Fig.\ \ref{fig:vdf_comp}. Recall that the $z$-independent model matches the two $z$-dependent models at $z=0$. As observed, at $z=1$ the three models are relatively similar to the $z=0$ curve, but their differences grow with redshift. At redshift $z=6$ the $z$-dependent model 2 predicts a orders of magnitude lower number density for higher dispersion velocities, which may be underestimated as suggested in \cite{Abe:2024rbf}. %For this reason, the $z$-dependent model 2 is considered to be a more conservative scenario.
Notice that while the $z$-dependent model 1 predicts a flatter distribution with redshift (i.e.\ more comparable galaxy number density at low and small velocity dispersions), the $z$-dependent model 2 predicts a steeper distribution with redshift (i.e.\ much higher galaxy number density at low rather than high velocity dispersion).  Thus, these two models follow different galaxy population trends found in the literature.

Finally, we emphasize that the $\phi(\sigma,z)$ distributions shown here describe the intrinsic galaxy population. Nonetheless, the galaxy lens distribution will be given by $\sigma^4\phi(\sigma,z)$, which incorporates the effect of the area of the Einstein radius, as given in the optical depth in Eq.\ (\ref{eq:tau3}).

\subsection{Lens Redshifts}\label{app:zL_dist}
%The lens redshift distribution in all the population models is shown in Fig.\ \ref{fig:zlzs2}. 
The percentage of lenses below the redshifts thresholds 1, 3, and 6 are shown in Table \ref{tab:zL_percent}. We find the three lens population models to have almost all events below $z=6$, yet there is a general trend that the $z$-dependent model 2 has more events at lower redshifts, compared to the other two models. 
\vspace{0.2cm}

\begin{table}[h!]
\centering
\begin{tabular}{|c|c|c|c|c|}
\hline
\textbf{Source Model} & \textbf{Lens Model} & \textbf{$z_L < 1$ }& $z_L < 3 $ & $z_L < 6$ \\
\hline
\hline
\multirow{3}{*}{Pop-III (K+16)} & $z$-ind & 25 & 80 & 98 \\
 & $z$-dep 1 & 29 & 84 & 98 \\
 & $z$-dep 2& 30 & 92 & 100 \\
\hline
\multirow{3}{*}{Q3-d (K+16)} & $z$-ind & 36 & 92 & 100 \\
 & $z$-dep 1 & 42 & 95 & 100 \\
 & $z$-dep 2 & 42& 97 & 100 \\
\hline
\multirow{3}{*}{Q3-nod (K+16)} & $z$-ind & 26 & 81 & 98 \\
 &$z$-dep 1 & 33 & 87 & 99 \\
 & $z$-dep 2 & 34 & 92 & 100 \\
\hline
\multirow{2}{*}{HS-nod-noSN } & $z$-ind & 22 & 75 & 96 \\
 & $z$-dep 1 & 27 & 81 & 97 \\
(B+20) &$z$-dep 2& 28 & 89 & 100 \\
\hline
\multirow{3}{*}{HS-nod-SN (B+20)} & $z$-ind & 22 & 76 & 96 \\
 & $z$-dep 1 & 21 & 74 & 95 \\
 &$z$-dep 2 & 28 & 89 & 100\\
\hline
\multirow{2}{*}{HS-nod-SN-high-accr } & $z$-ind & 23 & 77 & 96 \\
 & $z$-dep 1 & 29 & 82  & 97 \\
(B+20) & $z$-dep 2 & 30 & 90 & 100 \\
\hline
\end{tabular}
\caption{Simulation results showing distribution percentages for different $z$ thresholds.}
\label{tab:zL_percent}
\end{table}

\newpage 
\bibliographystyle{apsrev4-1}
\bibliography{References.bib}

\end{document}